\documentclass[11pt,twoside]{article}  
\usepackage{apn3conf}

\begin{document}   


\title{Multiple Point-Symmetric Ejections in IC\,4634}


\author{
Mart\'{\i}n A.\ Guerrero\altaffilmark{1,2},
Luis F.\ Miranda\altaffilmark{1}, 
You-Hua Chu\altaffilmark{2}}
\altaffiltext{1}{Instituto de Astrof\'{\i}sica de Andaluc\'{\i}a, CSIC, 
Apdo.\ Correos 3004, E-18080 Granada, Spain}
\altaffiltext{2}{University of Illinois, Department of Astronomy, 1002 W.\ 
Green St., Urbana, IL61801, USA}


\contact{Martin A. Guerrero}
\email{mar@iaa.es}


\paindex{Guerrero, M. A.}
\aindex{Miranda, L. F.}     
\aindex{Chu, Y.-H.}     


\authormark{Guerrero, Miranda, \& Chu}




\begin{abstract}          
We present a spatio-kinematical study of the planetary nebula (PN) 
IC\,4634 which has experienced several episodes of point-symmetric 
ejections oriented at different directions.  
The nebula displays two S-shaped low-ionization arcs that are probably 
related to two relatively recent point-symmetric ejections, the outer 
S-shaped arc representing a beautiful example of a bow-shock resolved 
in a PN.  
We report here the discovery of an arc-like string of knots at larger 
distances from IC\,4634 central star that represents a much earlier 
point-symmetric ejection.  
\end{abstract}


\section{Introduction}

IC\,4634 is one of the most spectacular point-symmetric planetary nebulae 
(PNe).  
Its remarkable double-S-shaped morphology is reflected in its kinematics, 
with two oppositely directed pairs of red- and blue-shifted features 
straddling the nebula (Schwarz 1993).  
The morphology and kinematics suggest a precession or rotation of the 
source that produced these highly symmetric outflows.  

We have used narrow-band archival {\it HST} images in the [O~{\sc iii}], 
H$\alpha$, and [N~{\sc ii}] lines to study the morphology of IC\,4634 and 
used complementary CTIO 4m telescope long-slit echelle observations of 
the H$\alpha$ and [N~{\sc ii}] $\lambda$6583 lines to carry out a detailed 
study of its spatio-kinematical structure.  

\section{Results}

The {\it HST} image of the central region of IC\,4634 (Figure~1-top-left) 
reveals several morphological constituents.  
The nebula has a sharp central elliptical shell surrounded by a 
misaligned elliptical envelope.  
Further emission found along PA$\sim135^\circ$ and $\sim315^\circ$ seems 
to form part of an irregularly shaped outer envelope.  
What makes IC\,4634 so remarkable are the two pairs of low-ionization, 
S-shaped arcs.  
In addition to these complex point-symmetric features, the image of 
the outermost regions of IC\,4634 (Figure~1-top-right) shows an 
arc-like string of knots located at $40^{\prime\prime}-50^{\prime\prime}$ 
north of the central star.  

The bottom pictures in Figure~1 reveals the bow-shock morphology of the 
outer North S-shaped arc (the South counterpart has similar morphology).  
The [O~{\sc iii}] and H$\alpha$ emission in these bow-shocks precede the 
[N~{\sc ii}] emission, indicating shock excitation.  
A shock velocity of 80-160~km~s$^{-1}$ is derived from the 
[O~{\sc iii}]/H$\alpha$ line ratio.  
The [N~{\sc ii}] emission from the bow-shocks shows developing 
instabilities. 
The [N~{\sc ii}]-bright string of knots arises from the edge of the 
bow-shocks, suggesting that they might be trailed material from the 
bow-shock.  

The H$\alpha$ and [N~{\sc ii}] echellograms provide further insights on 
the structure~of IC\,4634.  
The inner shell can be described as an ellipsoide with its northwestern tip 
receding from the observer.  
The inner low-ionization arcs expand at an almost constant V$_{\rm exp}$ of 
$\sim$20~km~s$^{-1}$ in opposite direction to the inner shell.
The radial velocity in the outer low-ionization arcs shows an abrupt 
change from the bow-shock structure to the linear string of knots, 
suggesting a deceleration.  
A counterpart of the outermost arc-like string of knots is detected to 
the south of IC\,4634 at a radial velocity similar to its systemic
velocity.

From these results, we conclude that IC\,4634 has experienced a complex 
history of multiple collimated outflows.    
These outflows have been ejected along different directions and very 
likely during different phases of the PN formation.  

\begin{figure}
\epsscale{0.84}
\plotone{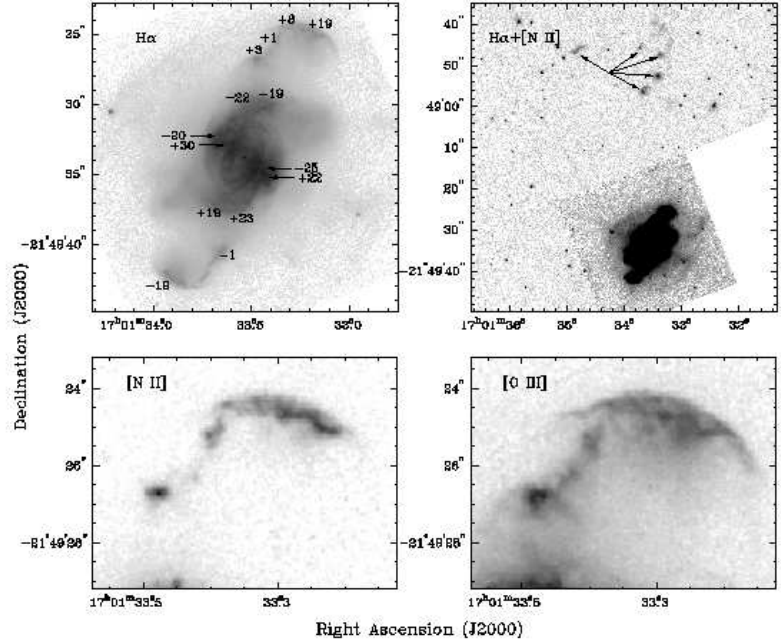}
\caption{
WFPC2 {\it HST} narrow-band images of IC\,4634.  
The radial velocity of different features of IC\,4634 (with respect to 
its systemic velocity) is shown on the H$\alpha$ picture (top-left).  
The location of the outermost string of knots is marked by arrows on 
the top-right picture.  
The  bottom pictures show close-ups in [N~{\sc ii}] and [O~{\sc iii}] 
of the North bow-shock.
}
\end{figure}

%


\end{document}